\documentclass[useAMS,usenatbib]{mnras}

\usepackage{graphicx}
\usepackage{scalefnt}
\usepackage{amssymb}
\usepackage{upgreek}
\usepackage{balance}

\newcommand{\yr}	    {\ifmmode \mathrm{yr} \else yr\fi}
\newcommand{\mpc}	{\ifmmode \,\mathrm{Mpc}^{-3} \else \,Mpc$^{-3}$\fi}
\newcommand{\Msun}	{\ifmmode \,\mathrm M_{\odot} \else $\,\mathrm M_{\odot}$\fi}
\newcommand{\Mhalo}	{\ifmmode M_{\mathrm{halo}} \else
  $M_{\mathrm{halo}}$\fi}
\newcommand{\Rvir}	{\ifmmode R_{\mathrm{vir}} \else $R_{\mathrm{vir}}$\fi}
\newcommand{\Moff}	{\ifmmode M_{\mathrm{off}} \else
  $M_{\mathrm{off}}$\fi}
\newcommand{\Toff}	{\ifmmode T_{\mathrm{off}} \else $T_{\mathrm{off}}$\fi}
\newcommand{\Mstar}	{\ifmmode {M}_{\star} \else ${M}_{\star}$\fi}

\title[Quenching vs. Quiescence]{Quenching vs. Quiescence: forming realistic massive ellipticals
  with a simple starvation model}
\author[T. A. Gutcke et al.]{Thales
  A. Gutcke$^1$\thanks{thales@mpia.de}, Andrea V. Macci\`o$^{2,1}$, Aaron
  A. Dutton$^2$, Greg S. Stinson$^1$
\\
$^{1}$Max-Planck-Institut f\"ur Astronomie, K\"onigstuhl 17, 69117 Heidelberg, Germany\\
$^{2}$New York University Abu Dhabi, PO Box 129188, Saadiyat Island, Abu Dhabi, United Arab Emirates}

\begin{document}

\pagerange{\pageref{firstpage}--\pageref{lastpage}} \pubyear{---}

\maketitle

\label{firstpage}

\begin{abstract}
The decrease in star formation (SF) and the morphological change
necessary to produce the $z=0$ elliptical galaxy population are
commonly ascribed to a sudden quenching event, which is able to rid the central galaxy of its cold gas
reservoir in a
short time. Following this event, the galaxy is able to prevent further
SF and stay quiescent via a maintenance mode.  We test
whether such a quenching event is truly necessary using a simple model of
quiescence. In this model, hot gas (all gas above a
temperature threshold) in a $\sim 10^{12} \Msun$ halo mass galaxy at
redshift $z\sim 3$ is prevented from cooling. The cool gas continues
to form stars at a decreasing rate and the galaxy stellar mass,
morphology, velocity dispersion and position on the color magnitude
diagram (CMD) proceed to evolve. By $z=0$, the halo mass has grown to
$10^{13} \Msun$ and the galaxy has attained {characteristics typical of
an observed} $z=0$ elliptical galaxy.  Our model
is run in the framework of a cosmological, smooth particle
hydrodynamic code which includes SF, early stellar feedback, supernova
feedback, metal cooling and metal diffusion.  Additionally, we
post-process our simulations with a radiative transfer code to create a mock CMD. In contrast to previous assumptions
that a pure ``fade away'' model evolves too slowly to account for the
sparsity of galaxies in the ``green valley'', we demonstrate crossing
times of $\lesssim 1$\,Gyr. We conclude that no sudden quenching event
is necessary to produce such rapid colour transitions.

\end{abstract}

\begin{keywords}
\end{keywords}

\section{Introduction}


The existence of red elliptical galaxies has perplexed galaxy
formation and evolution modelers since the first observations of the
bi-modality in the color-magnitude diagram (CMD) by
\citet{Takamiya1995}.  Large-scale surveys such as the Sloan Digital
Sky Survey (SDSS) increased the statistical sample and broadened the
knowledge of this bi-modality \citep{Strateva2001, Baldry2004,
  Baldry2006}. 
Before then, the simple morphology and surface brightness profiles of
elliptical galaxies lead to the belief that these galaxies were an
early stage in the evolution process of galaxies, leading to their
commonly used name early-type galaxy. 


Further strengthening the distinction of the two  populations,
research showed bi-modality both locally and at high redshift, not
only in galaxy colors and morphologies but also their star formation
rates (SFRs) \citep[e.g.][]{Bell2004, Brammer2009}.
\citet{Kauffmann2003} found that red galaxies have on average older
stellar populations, higher surface stellar mass densities, and
dominate at stellar masses above $10^{10.5}\Msun$.

A deeper understanding of stellar evolution has led to the insight
that the red color originates from the old stellar
populations. Meaning that many {``intrinsically'' red} galaxies are at a later
stage in galaxy evolution. Blue galaxies necessarily migrate to the
red sequence through a cessation of star formation \citep{Bell2004,
  Faber2007}.  The resulting gap between the blue galaxies and the red
sequence in the CMD is termed the ``green valley''. The galaxies in
the green valley constitute a transiting population between the two
regions \citep[e.g.][]{Bell2004, Faber2007,Martin2007,
  Schiminovich2007, Wyder2007, Mendez2011, Goncalves2012}. Various
  authors have defined the green valley differently, some
  specifying a region in the UV-optical diagram, some using the
  optical-optical diagram. Although the precise
  definition can vary, the results seem to stay consistent.

The more rapid the transition through the green valley, the sparser
this region in the CMD will be. But \citet{Peng2010} found the
sparsity consistent with an overlap of a fast and a slow quenching
mechanism. The timescale commonly used to distinguish fast and slow
modes lies around $1$\,Gyr \citep{Schawinski2007, Salim2014}. But
\citet{Martin2007} recognised that the number density in the green
valley does not deliver a crossing time by itself. Using H$\delta_{\rm
  A}$
and D$_{\rm n}$(4000) spectral indices and physically reasonable star formation
histories, they found that quenching timescales vary widely and green
valley crossing times can be around and above $2.5$\,Gyr for some
galaxies. \citet{Yesuf2014} used SDSS, GALEX and WISE data to estimate
a transit time of $7$\,Gyr for the slowest galaxies, assuming that by
far the majority ($\sim 74\%$) of red sequence galaxies follow the
slow quenching mode. {\citet{Trayford2016} used the EAGLE simulation to
estimate crossing times. Although they distinguish between three
characteristic evolutionary paths across the green valley, they find no distinct
difference in crossing times between them. Using the Galaxy Zoo
morphological classification, \citet{Smethurst2015} identified three
seemingly distinct quenching timescales: morphologically smooth
galaxies quench rapidly, attributed to major mergers. Disc-like
galaxies quench
on slow timescales caused by secular evolution. Their intermediate
population reddens in intermediate timescales, which they attribute to minor mergers and galaxy interactions.}

The now commonly used technique to match the luminosity derived
stellar masses of galaxies with a dark matter halo mass (generally
termed ``abundance matching'') gives rise to a clearer picture of how
well and efficiently a galaxy at a given dark matter halo mass turns
available gas into stars. At halo masses below and above $10^{12}
\Msun$, the star formation efficiency (SFE) decreases
significantly. The peak efficiency around $10^{12} \Msun$ is of the
order $20\%$ compared to the cosmic baryon fraction. Why star
formation is inefficient at high halo masses ($>10^{13} \Msun$) is a
key question in understanding the existence and formation of
``red-and-dead'' galaxies. 

{Galaxy formation models have had trouble simulating the
  bi-modality of galaxies and, in general, struggle with explaining the
  low efficiency of star formation especially at the high mass end
  \citep[e.g.][]{Benson2003}. In attempts at} modelling the formation of elliptical galaxies, theorists have
proposed and tested a large array of possible mechanisms to quench
star formation and trigger a morphological change. In their various
implementations, any of these could be a viable physical mechanism
preventing the cooling in our quiescence model.
The clustering of active galactic nuclei (AGN) host galaxies in the
green valley suggest a role for AGN feedback
\citep[e.g.][]{Nandra2007, Hasinger2008, Silverman2008, Cimatti2013}. 

{In attemps to follow what observations seem to show and ameliorate the
  over-production of stars, many models} implemented both the slow and fast
quenching mechanisms by assuming that active galactic nuclei (AGN) quenching (fast) involves
episodic, and possibly violent, feedback, while halo quenching (slow)
involves the suppression of gas cooling. These models broadly
reproduce the colors and mass functions of present day galaxies
\citep[e.g.][]{Croton2006, Cattaneo2006, Schaye2015}.

Semi-analytical models and hydrodynamical simulations began
implementing one mode of AGN feedback
\citep[e.g.][]{DiMatteo2005,Croton2006, Bower2006}.  Based on
observations of two different ways in which  black holes (BHs)  effect
surrounding gas, later models distinguished ``radio mode'' and ``quasar
mode'' feedback (\citealt{Sijacki2007, Somerville2008, Fanidakis2011},
and see also \citealt{Gutcke2015} for some effects of the two mode
model). The former is induced when BH accretion rates are low, while
the latter is active when mergers or disk instabilities trigger high
accretion rates. These models successfully create a bi-modal galaxy
population with quenched, red galaxies.

But there is still no consensus whether one or two modes are
necessary, as shown by recent large-box cosmological simulations such
a Illustris \citet{Vogelsberger2014} that employs a two mode model and
the Eagle simulations suite \citet{Schaye2015} which uses one mode of
AGN feedback.  And yet the questions of whether and how the energy
released by the AGN can couple efficiently to the surrounding gas is
still unresolved \citep[e.g. see][]{Cielo2014}
%

At lower galactic masses, feedback from supernovae and stellar winds
\citep[e.g.][]{DekelSilk1986, Murray2005, Stinson2013} effectively
decreases the amount of gas available in the centers of galaxies and
thus forces star formation to dwindle. 
Another viable quenching scenario is morphological quenching in which
a stellar spheroid helps the gas disk to become stable against
fragmentation, which leads to less star formation \citep{Martig2009,
  Martig2013, Genzel2014, Forbes2014}.

The long-term suppression of the external gas supply once the halo
mass grows above a threshold mass of $\sim10^{12}\Msun$ also
counteracts excessive star formation. This can happen either via
virial shock heating \citep{BirnboimDekel2003, Keres2005,
  DekelBirnboim2006, Keres2009}, or by gravitational infall heating
\citep{DekelBirnboim2008, BirnboimDekel2011, KhochfarOstriker2008},
which can be aided by AGN feedback coupled to the hot halo gas
\citep{DekelBirnboim2006, Cattaneo2009, Fabian2012}.
The incidence of cosmic rays has been largely neglected in most galaxy
evolution models. These energetic particles have the potential to
increase the pressure of the gas, retaining energy and preventing it
from being radiated away. Increased scale heights of the disc can
prevent further star formation \citep{Pfrommer2016}. 

Although investigation into quenching mechanisms has gone far and
wide, no general agreement has emerged as to which effect dominates in
massive galaxies. This has led to models such as \citet{Dutton2015}
and \citet{Gabor2015} who attempt to understand the effects of
quenching without defining a physical origin.  Motivated by the
success of halo quenching models \citep{Cattaneo2006},
\citet{Dutton2015} investigated a `forced quenching' scenario where
cooling and star formation are shut off at $z\sim 2$. They show that
this results in present day elliptical galaxies with stellar masses
and structural properties in broad agreement with observations.

This paper extends the work of \citet{Dutton2015} by constructing a
quiescence model that allows some cooling and star formation to occur
after the halo quenching begins. We present the characteristics of the
resulting elliptical galaxies and quantify the deviations from
observations. Our model makes no assumptions about the cause of the
limited cooling and can be compared to all of the above quenching
mechanisms.

The structure of this paper is as follows: In section \ref{sec:model}
we describe our quiescence model in detail. In section \ref{sec:obs}
we compare our resulting galaxies to observations: in \S\ref{sec:msmh}
the stellar mass-halo mass relation; in \S\ref{sec:sfh} the star
formation histories (SFHs); in \S\ref{sec:ssfr} the specific star
formation rate (sSFR) - stellar mass relation; in \S\ref{sec:cmd} the
CMD; in \S\ref{sec:rm} the size-mass relation and in \S\ref{sec:vm}
the Faber-Jackson relation. Section \ref{sec:summary} discusses and summarizes 
our results.

\begin{table*}
  \centering
  \begin{tabular}{l|c|c|c|c|c|c|c|c|c|c|c}
\hline
Name & $T_{\rm off}$ & $M_{\rm off}$ & $M_{\rm vir}$  & $M_{\star}$
    & $R_{\rm vir}$  & $m_{\rm dark}$ & $m_{\rm gas}$
& $\epsilon_{\rm dark}$ & $\epsilon_{\rm gas}$ & $N_{\rm dark}$ & $N_{\rm gas}$\\
& [$10^5$K] & [$10^{12}$ M$_{\odot}$] & [$10^{13} $M$_{\odot}$] &
                                                                  [$10^{11}
                                                                  $M$_{\odot}$]
    & [kpc] & [$10^6 $M$_{\odot}$] & [$10^6 $M$_{\odot}$]
& [kpc] &[kpc] & million & million\\
\hline
Halo1       & 1.0 & $2.48$ & 1.375 & 0.85 & 510 & 20.71 & 1.139 & 2.127 & 1.192 & 1.24 & 1.25 \\
Halo2       & 1.0 & $2.37$ & 1.222 &1.55 & 509 & 20.71 & 1.146 & 2.033 & 1.114 & 1.26 & 1.04 \\ 
Halo3       & 1.0 & $1.94$ & 1.229 & 1.17 & 496 & 20.71 & 1.146 & 1.987 & 1.114
                                               & 1.34 & 1.14 \\
Halo3late       & 1.0 & $4.11$ & 1.268 & 2.62 & 495 & 20.71 & 1.145 & 2.127 & 1.192 & 1.34 & 1.30 \\
Halo2h & 1.0 & $2.39$ & 1.232 & 1.70 & 490 & 2.6   & 0.140 & 1.060 & 0.600 & 15.4 & 10.05\\
\hline
Halo1c       & - & - & 1.363 & 6.74 & 507 & 20.71 & 1.144 & 2.127 & 1.192 & 1.26 & 1.11 \\
Halo2c       & - & - & 1.241 & 8.68 & 492 & 20.71 & 1.147 & 2.127 & 1.192 & 1.26 & 1.07 \\ 
Halo3c       & - & - & 1.268 & 7.5 & 495 & 20.71 & 1.147 & 2.127 & 1.192 & 1.34 & 1.18 \\
\hline
H3-5e4   & 0.5 & $1.02$ & 1.259 & 0.45 &  472 & 165.7 & 9.509 & 3.989 & 1.709 & 0.17 & 0.17\\
H3-1e5P   & 1.0  & $1.39$ & 1.268 & 1.09 & 471 & 165.7 & 9.416 & 3.989 & 1.709
                                               & 0.17 & 0.17\\
H3-1e5   & 1.0  & $1.02$ & 1.249 & 0.63 & 471 & 165.7 & 9.416 & 3.989 & 1.709 & 0.17 & 0.17\\
H3-5e5   & 5.0  & $1.02$ & 1.223 & 1.11 & 468 & 165.7 & 9.453 & 3.989 & 1.709 & 0.17 & 0.17\\
H3-1e6   & 10.0 & $1.02$ & 1.219 & 1.54 & 467 & 165.7 & 9.262 & 3.989 & 1.709 & 0.17 & 0.17\\
\hline
\end{tabular}
\caption{Simulations run for this analysis and their respective
  parameters: $T_{\rm off}$, the temperature above which gas cooling
  was turned off; $M_{\rm vir}$, the virialised mass at $z=0$;
 {$M_{\star}$, the stellar mass inside 0.1\Rvir~ at $z=0$; $R_{\rm
    vir}$, the radius at $z=0$ enclosing 200 times the critical density;} $m_{\rm
    dark}$, the mass of a dark matter particle in the zoom-in region;
  $m_{\rm gas}$, the mass of a gas particle in the zoom-in region;
  $\epsilon_{\rm dark}$, the dark matter particle softening length;
  $\epsilon_{\rm gas}$, the minimum gas particle smoothing length; $N_{\rm
    dark}$, the number of dark matter particles in the zoom-in region;
  $N_{\rm gas}$, the number of gas particles in the simulation. {The
  table is divided into three sections: The first five simultions are the
ones we will focus on most in the analysis. The second group of three
are the control simulations for our three initial conditions and the
third group of five are low resolution simulations we ran to test
various parameter choices.}}
\label{tab:sims}
\end{table*}
%
%
\section{Numerical Simulations}
\label{sec:model}

Numerical simulations have been performed  using a improved version of
the smooth  particle hydrodynamics (SPH) code  {\sc gasoline}
\citep{Wadsley2004,  Keller2014},  this new version  promotes mixing
and strongly  alleviates the  well known issues  of the  classical SPH
formulation \citep{Agertz2007}.
The  baryonic physics  treatment is  the  same adopted  for the NIHAO
simulation suite \citep{Wang2015}. Namely the code includes a subgrid
model for metals mixing \citep{Wadsley2008}, ultraviolet heating and
ionization and cooling due to  hydrogen, helium and metals as detailed
in \citet{Shen2010}.

The star formation  and stellar feedback follow  the implementation of
the MaGICC  simulations by  \citet{Stinson2013}. We set the star
formation threshold as suggested in \citet{Wang2015}, using a critical density of 5.17 cm$^{-3}$. The density threshold is
defined as a kernel of 52 gas particles within a sphere of radius
equal to the softening length. Stars can return
energy back to the inter-stellar medium (ISM) via blast-wave supernova
(SN) feedback  \citep{Stinson2006} and  via ionizing  radiation from
massive stars (early stellar feedback) before they turn in SN
\citep{Stinson2013}.  Metals are produced by  type II and type  Ia SN.
Stars also return  mass to the  ISM via  asymptotic giant branch
stars.  We adopt a \citet{Chabrier2003} stellar initial mass function
which sets the fraction of stellar mass that results in SN and winds.

We select  three haloes with a {present-day} mass of about $10^{13}\,
\Msun$ from the Planck simulations  from \citet{Dutton2014} to be
re-run at higher resolution. The initial  conditions were  created
using  a modified  version of  {\sc grafic2}  \citep{Penzo2014}; the
starting redshift  is $z_{\rm start}=99$.

We run three halos which we call Halo1, Halo2 and Halo3. Halo1 has a quiescent merger
history, while Halo2 and Halo3 have one major merger each (at 11 Gyr
and 7 Gyr, respectively) and some minor mergers in their past. We also
run a control simulation of each halo without any changes to the
cooling (Halo1c, Halo2c, Halo3c). To
check for convergence, we run Halo2 at three
different resolution levels and Halo3 at two.  See table \ref{tab:sims} for
an overview of the simulations and the halo properties.

We use the Amiga Halo Finder (AHF, \citealt{Knollmann2009}) to identify
gravitationally bound structures and to track these throughout cosmic
time.  We rely on the AHF output for estimating the viral radius,
which is defined so that it encloses an overdensity of 200 times the
critical density of the Universe. 


\begin{figure}
  \centering
 \includegraphics[width=0.48\textwidth]{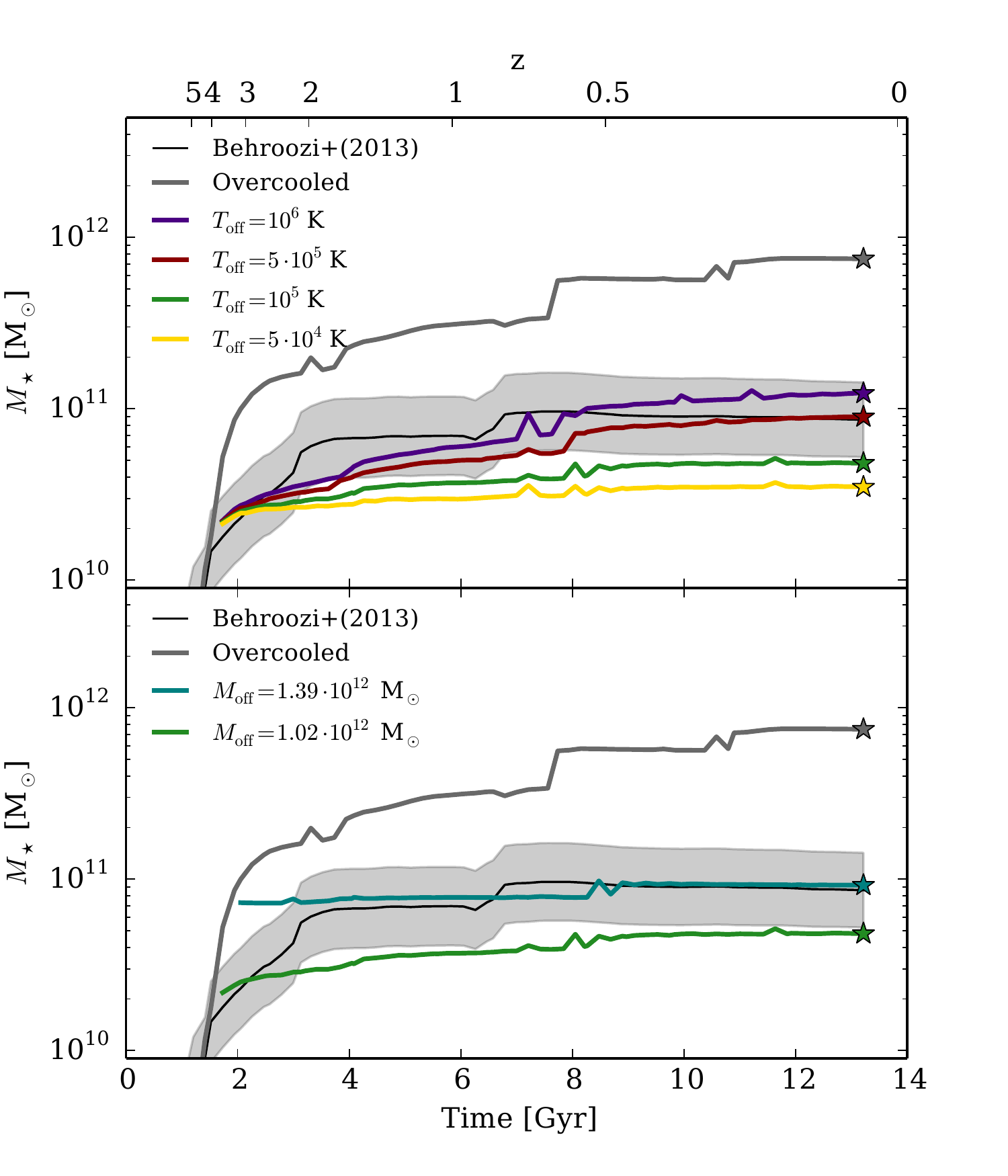}
  \caption{{\it Top:} Halo3 simulated 5 times with varying parameter
    $T_{\rm off}$ and $M_{\rm off}=1.02\times10^{12}\Msun$ (Halo3c,
    H3-5e4, H3-1e5, H3-5e5, H3-1e6). {\it
      Bottom:} Halo3 with varying parameter $M_{\rm off}$ and $T_{\rm
      off}=10^5$K (Halo3c, H3-1e5P, H3-1e5). {The stellar mass is
      defined as the sum of all stellar particle masses within 0.1
      \Rvir.}}
  \label{fig:toff}
\end{figure}

\begin{figure*}
  \centering
 \includegraphics[width=0.98\textwidth]{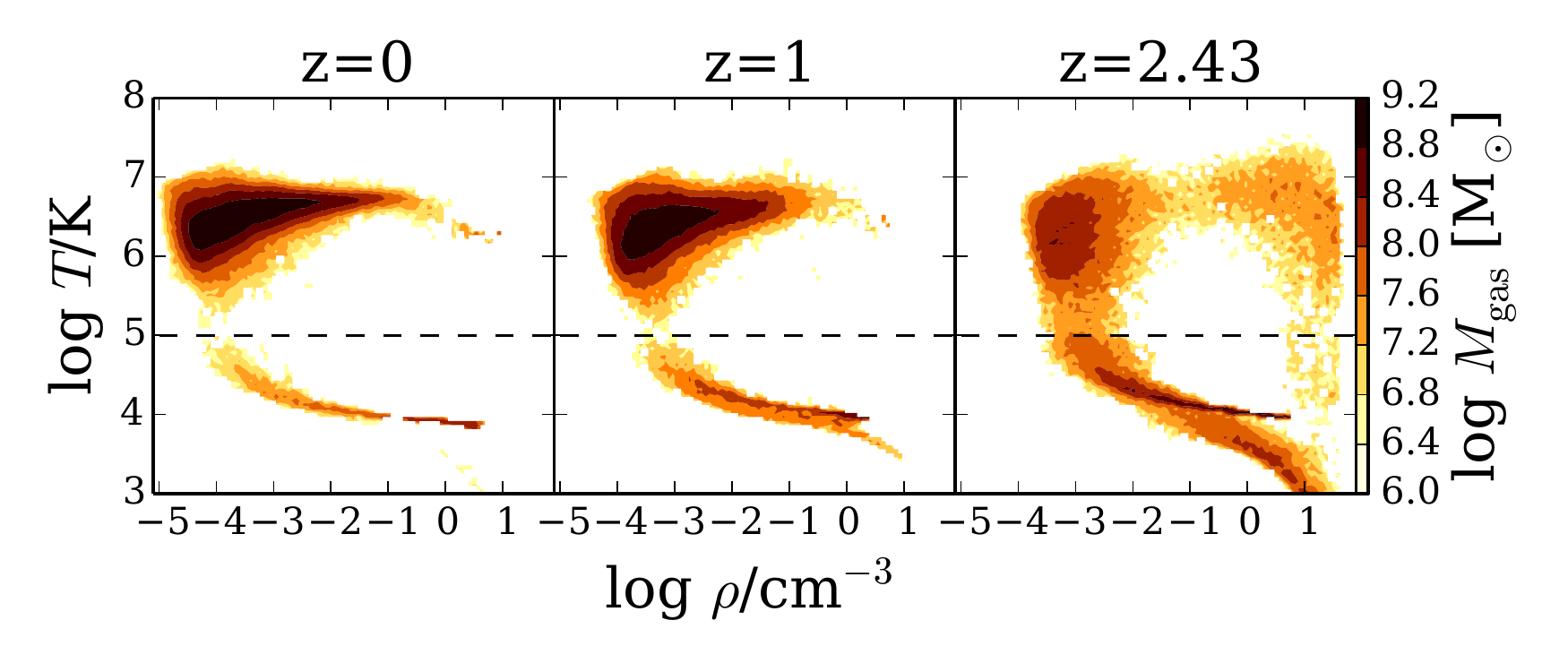}
  \caption{Phase diagrams for Halo3 at three different redshifts,
    $z=2.43$ (where the cooling mode is switched), $z=1$ and $z=0$. Bin
    widths are log($\rho/[{\rm cm}^{-3}]) = 0.07$ and log$(T/[{\rm
        K}]) = 0.05$.  The dashed line indicates the $T_{\rm
      off}=10^{5}$K. The parameter forces the disk and hot halo to
    separate at this temperature. }
  \label{fig:phase}
\end{figure*}

\subsection{A simple model of quiescence}

 Numerical simulations without any form of feedback
consistently overestimate the number of stars created in low mass and
high mass galaxies. At the low mass end, it has become accepted to
implement feedback from supernova and even radiation from young
stars. For high mass galaxies often AGN
feedback is assumed to limit the cooling. But observations are still
very ambivalent as to whether this heating mechanism truly is able to
affect such a large fraction of the gas in the centers of galaxies
(see \citealt{Cielo2014} for numerical simulations showing this). Many
other physical mechanisms have been tested in simulations to limit SF,
each varying in physical viability and success. 

Here we are interested in a much simpler question: would it be enough
to prevent accretion of cold gas to create a realistic elliptical galaxy?
In other words, can we create a red-and-dead galaxy simply by starving
galaxies above a critical mass?
To test this, we implement a very simple model to force
a quiescent evolution: up to a critical total mass (called \Moff)
a galaxy is allowed to evolve freely. Then, once this mass is reached
we shut off cooling for all gas particles above a critical temperature
\Toff. The critical mass and temperature are the only two parameters
of our simple model.
It is important to notice that even for haloes above $\Moff$ we
halt neither cooling nor star formation completely.
Cold gas (i.e. below \Toff) is still
able to cool and form stars, which in turn will be sources of stellar feedback.
We simply prevent the cooling (and hence accretion)
of gas above $\Toff$ onto the galaxy.

%
In figure \ref{fig:toff} we show the effects of different choices for the
two parameters in our model on the evolution of the stellar mass
for Halo3 at low resolution ({third} part of Table
\ref{tab:sims}). {Here and in the rest of this work, the stellar mass
is defined as all stellar particle masses within 0.1\Rvir.}
The black line shows the expected evolution of the stellar mass according
to  abundance matching results from \citet{Behroozi2013}, the (thick)
dark grey
line represents the simulation without suppressed cooling which
overproduces the final stellar mass by a factor
of almost 10.
In the upper panel we present results for four different values of $\Toff$ at a fixed
$\Moff=1.02\times10^{12}\,\Msun$, while in the lower panel we vary $\Moff$
at a fixed $\Toff=10^5$\,K.

{This simple study seems to justify using
$\Moff=1.4\times10^{12}\,\Msun$. However, increased resolution slightly
changes the star formation history \citep[due to lower stellar mass, see][]{Mayer2008} and we have to increase
$\Moff$ to $2\times10^{12}\,\Msun$ to get the correct final stellar
mass. So, for the rest of this work, the fiducial parameters for
our model are: $\Toff=10^5$\,K and $\Moff=2\times10^{12}\,\Msun$.
There is some variation from simulation to simulation in the parameter \Moff, given that we restart from the snapshot with a halo mass closest to the fiducial $\Moff$ value, and snapshots are on average 200 Myr apart.}
Given the intrinsic simplicity of our model we didn't see the need for
any more fine tuning of our parameters.


%
Fig. \ref{fig:phase} shows the phase diagram for Halo3 at three
redshifts, $z=2.43$, $1$, and $0$. The black dashed line indicates our
choice of \Toff. This choice is in line with previous studies
such as \citet{Gabor2015}. They heat the circum-galactic gas to
$10^{5.4}$\,K, preventing further star formation and successfully
producing hot halos. The temperature is a local minimum in the cooling
curve of simulations, distinguishing free-free emission and helium
cooling. The temperature break forces the Halo3 phase
diagram to separate along this temperature, resulting in a hot halo
phase in the upper left and a denser, colder disk phase below our
temperature threshold. The disc phase ($T\lesssim10^4$\,K,
$\rho\gtrsim10^{-2}$\,cm$^{-3}$) is depleted over time since each
cycle of star formation produces stellar winds and supernovae that
heat a fraction of the gas to above the threshold temperature, thus
making it unavailable for further cooling and star formation.

\newcommand{\mysize}{0.16}
\begin{figure*}
  \centering
 \includegraphics[width=\mysize\textwidth]{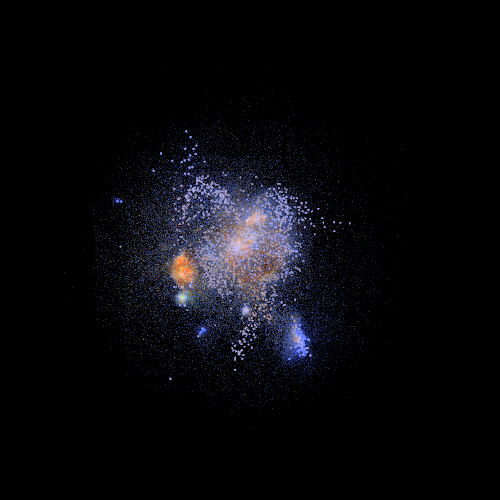}
\includegraphics[width=\mysize\textwidth]{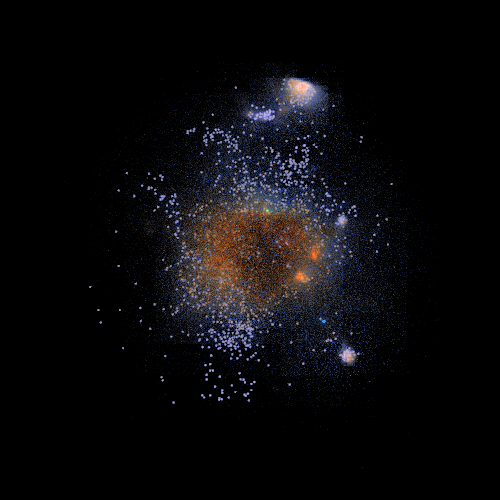}
\includegraphics[width=\mysize\textwidth]{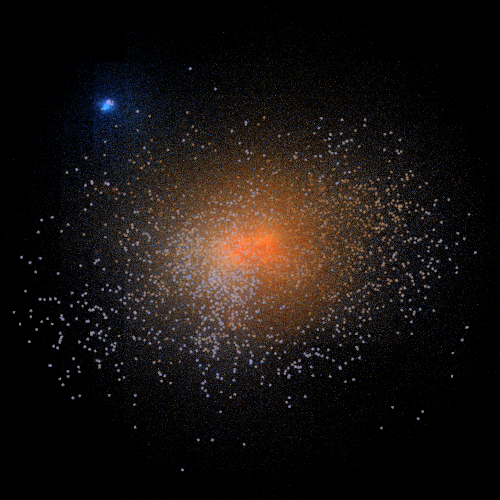}
\includegraphics[width=\mysize\textwidth]{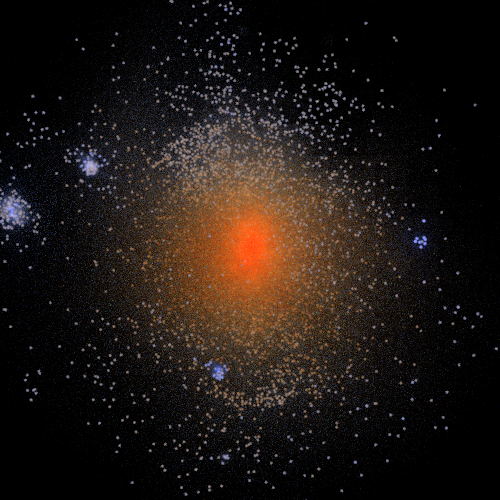}
\includegraphics[width=\mysize\textwidth]{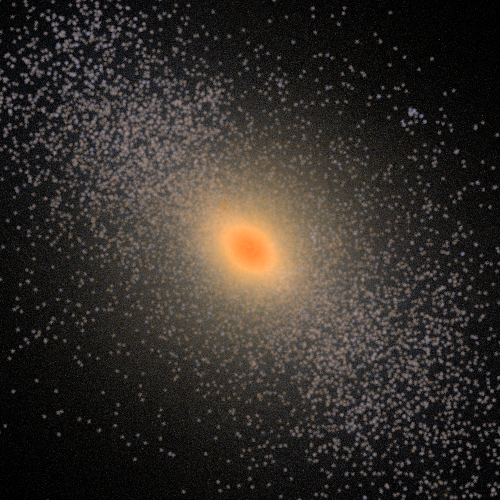}
\includegraphics[width=\mysize\textwidth]{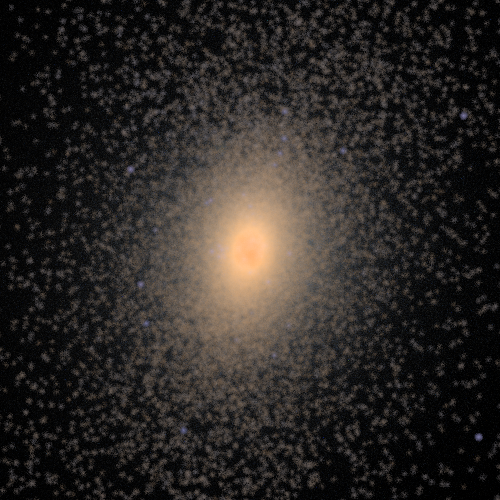}
  \caption{$50\,$kpc$\times50\,$kpc \texttt{SUNRISE} RGB images of
    Halo1 at redshifts (Gigayears) 2.9 (2.2), 2.6 (2.5), 1.8 (3.5),
    1.6 (4.0), 0.7 (7.1), and 0.0 (13.8), from left to right. The
    onset of the quiesence mode is at 2.2\,Gyr.}
  \label{fig:sunrise}
\end{figure*}

In this analysis, we are interested in affecting the gas and creating
a galaxy that has a morphology and other {properties} typical of the
elliptical population. This allows us to get a grasp on the conditions
necessary in simulations for creating a quiescent environment.  Our
primary concern is to understand whether a pure fade-away model where
gas is slowly consumed by star formation can match observational
constraints without an abrupt, violent quenching event. {The
quiescence model makes no assumptions about the
physical process that quenches a galaxy. Being agnostic about the mechanism can be used to our
advantage.} It can in principle be either
a model of inefficient cooling or of a supplementary heating mechanism
that counteracts cooling. This way, our results can be freely compared
to a variety of quenching models.
%
%
%
\section{Comparing to observations}
\label{sec:obs}
We compare our model galaxies with observations of $z=0$ elliptical
galaxies. We compare the stellar mass-halo mass relation, the SF
history, the size-mass relation, the stellar mass-velocity dispersion
relation and the transition from the blue cloud to the red sequence on
the color magnitude diagram. {To obtain a fairer comparison
with observables, we run the Monte Carlo radiative transfer code \texttt{SUNRISE}
\citep[v5.0,][]{Jonsson2006}.
The code uses the stellar population
synthesis model Starburst99 \citep{Leitherer1999}, assumes a Kroupa
initial mass function (IMF) and traces 10 million rays par galaxy. The
calculation includes dust estimates that absorb and scatter the rays. The dust mass is set to 0.4 times the
metal mass for each gas particle. Additionally, dust within the birth cloud around newly formed star particles
($t_{\rm form} < 10$\,Myr) is modeled in sub-grid using
\texttt{MAPPINGS III} \citep{Groves2008}. 

For the transformation from the particle positions to
the grid needed by \texttt{SUNRISE}, the radius of a gas particle
is defined as the distance to the 32nd nearest neighbour. The stellar radius is defined
as 0.2 times the gravitational force softening.
Figure \ref{fig:sunrise} shows six \texttt{SUNRISE} images of the
Halo1. From left to right the redshifts (Gigayears) are 2.9 (2.2),
2.6 (2.5), 1.8 (3.5), 1.6 (4.0), 0.7 (7.1) and 0.0 (13.8)}
Morphologically, the galaxies look elliptical and red.
%
\begin{figure}
  \centering
 \includegraphics[width=0.48\textwidth]{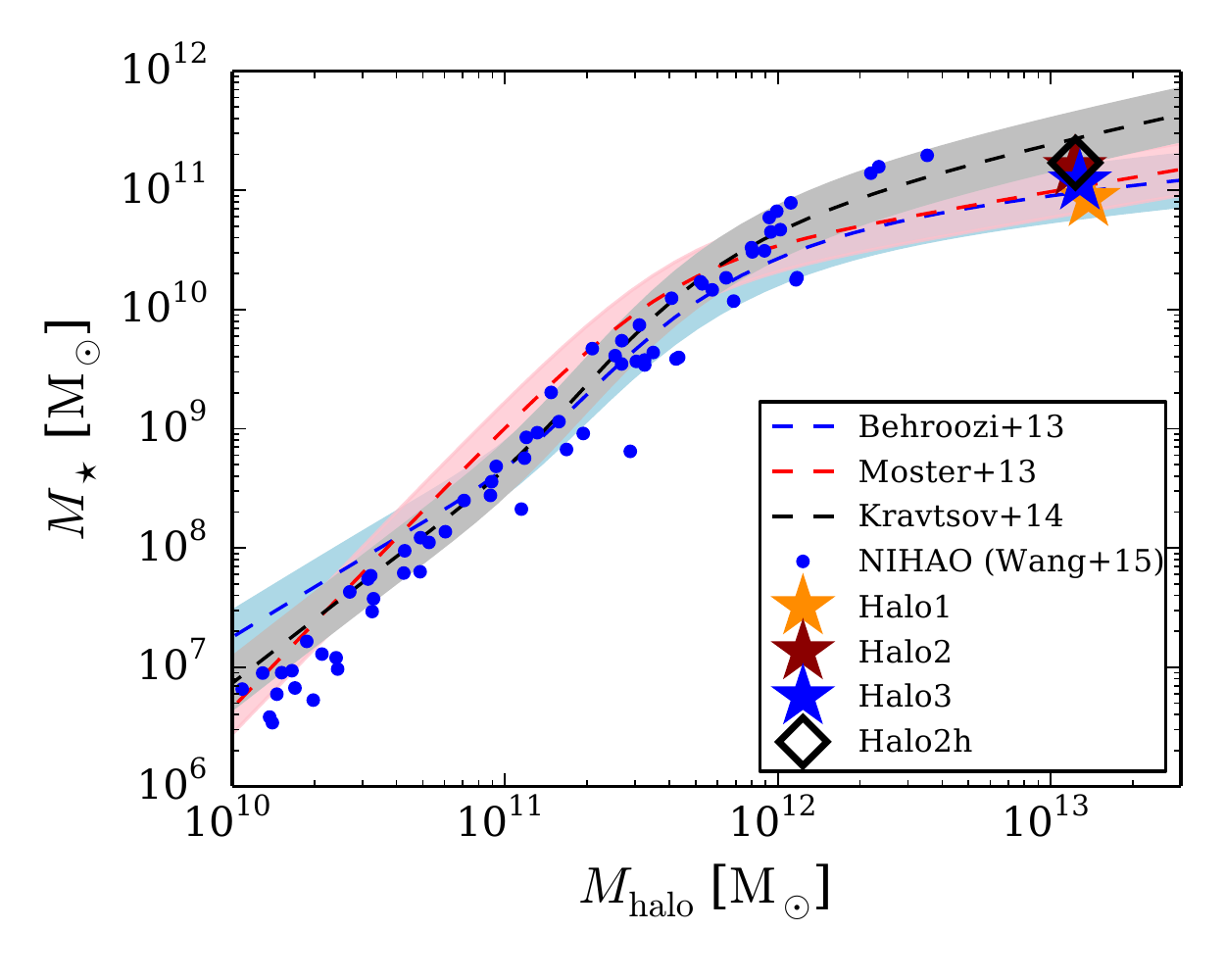}
  \caption{$z=0$ Stellar mass - halo mass relation for our galaxies
    (stars and diamond) and those of the NIHAO sample (blue dots),
    which use the same physics until we switch into quiescence
    mode. The blue dashed line and band shows observational data based
    on abundance matching from \citet{Behroozi2013}. Red line with a
    pink band shows the results of \citet{Moster2013} and black line
    with grey band shows results of \citet{Kravtsov2014}.}
  \label{fig:nihao}
\end{figure}

\subsection{Stellar mass - halo mass relation}
\label{sec:msmh}
The relation between the stellar mass of a galaxy and the total mass
of the halo including dark matter is a strong constraint on galaxy
formation models. For the stellar mass, \Mstar, we consider all stars
within $10\%$ of the virial radius, \Rvir. The total halo mass,
\Mhalo, includes all particles within \Rvir. The relation constructed
between these two quantities is our primary observational constraint,
and we choose our two model parameters to reproduce
it. Figure~\ref{fig:nihao} shows the abundance matching results for
\Mhalo-$\Mstar$ from \citet{Moster2013}, \citet{Behroozi2013} and
\citet{Kravtsov2014}. The star markers show our three elliptical halos
at $z=0$ and the blue dots show the high resolution NIHAO sample. The
NIHAO sample follows the relation across two orders of magnitude in
halo mass, from $10^{10} - 10^{12}\,\Msun$. Our quiescence model
galaxies also lie on the relation at $10^{13}\Mhalo$. In this context,
our model is the high mass extension of the NIHAO sample.

\begin{figure}
  \centerline{
 \includegraphics[width=0.47\textwidth]{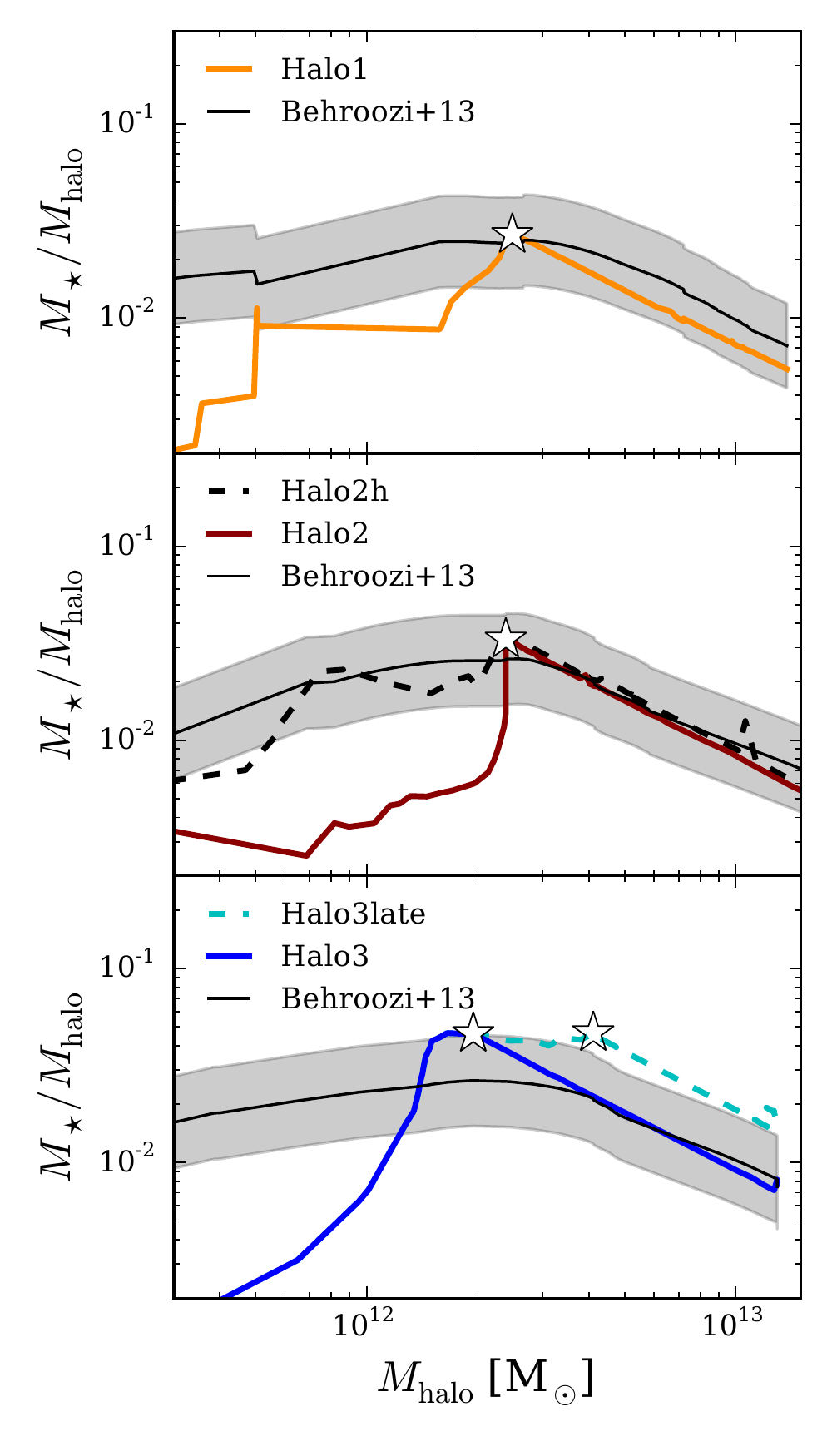}}
  \caption{Stellar mass/halo mass - halo mass relation for our galaxies. The
    gray band shows observational data based on abundance matching
    from \citet{Behroozi2013}. The colored bands show the $1-\sigma$ scatter
for the data.}
  \label{fig:ms-mh}
\end{figure}
Figure \ref{fig:ms-mh} shows the evolution of \Mhalo-\Mstar/\Mhalo~of
our three galaxies (one galaxy per panel) at the fiducial resolution
level. The black stars mark the point at which the cooling mode is
switched from regular metal-line cooling to the quiescence mode. The
black solid lines show the abundance matching results reported by
\citet{Behroozi2013}, where the gray band encloses the $1-\sigma$
scatter.  We match the slope of the relation after the onset of the
quiescence mode by construction, since the parameter $\Toff$
determines its behaviour. A lower $\Toff$ produces a {steeper}
slope, while a higher $\Toff$ produces a flatter evolution. The
parameter $\Moff$ was also chosen to match this relation since
the knee generally appears around $2\times10^{12}\Msun$.  Halo3 in the
lowest panel of figure \ref{fig:ms-mh} shows a slight bump above the
abundance matching area around $10^{12}\;\Msun$ due to a merger.
{We also show the relation for Halo3late, which is the same as
  Halo3 except that $\Moff$ is higher by a factor of about two. This
  leads to a higher final stellar mass. The change in final stellar
  mass and the intrinsic scatter in the \citet{Behroozi2013} relation
  show that a range of $\Moff$ are allowed while still reproducing the
  necessary shape.}

\begin{figure}
  \centering
 \includegraphics[width=0.47\textwidth]{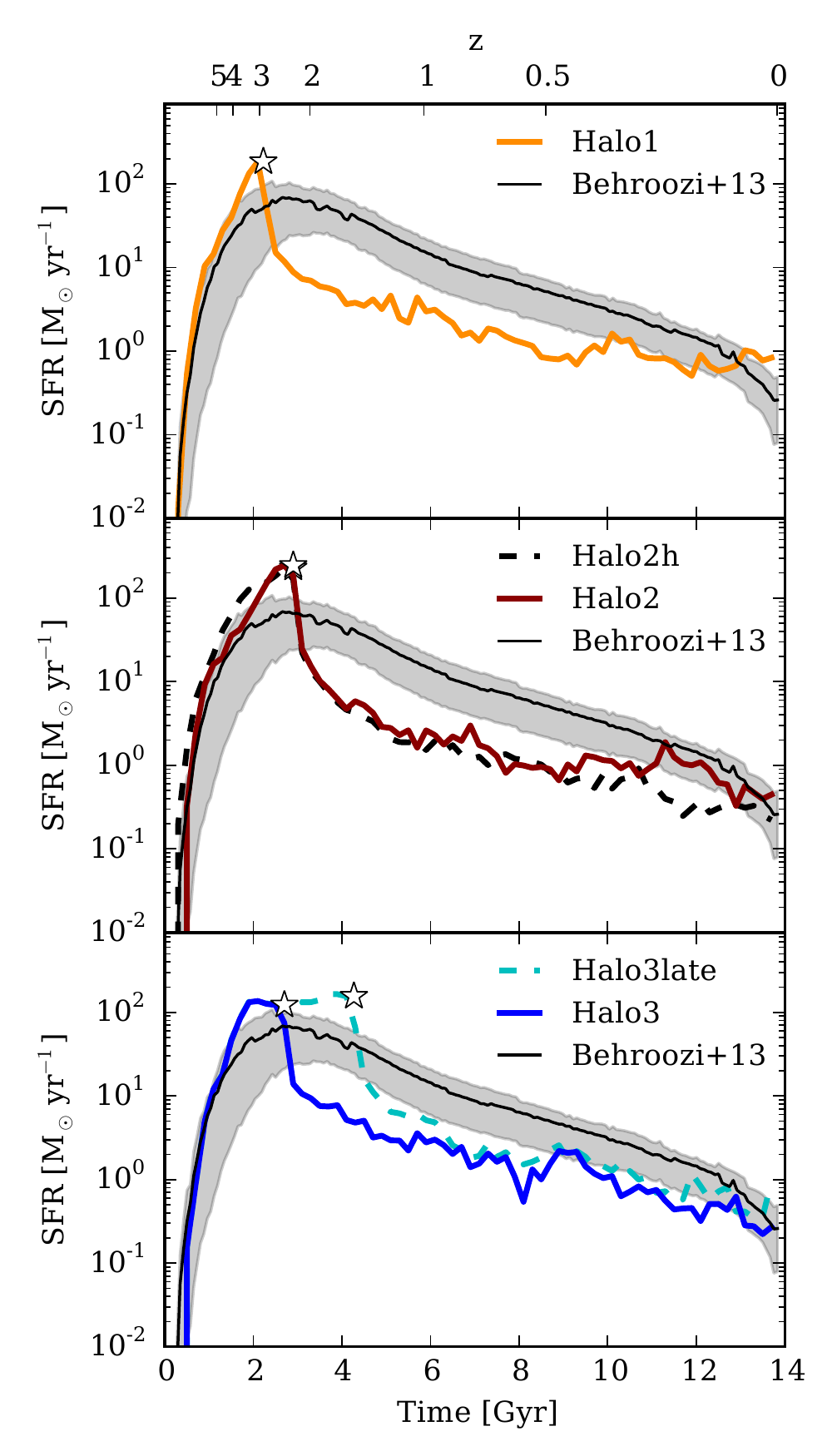}
  \caption{The star formation history of our galaxies, with the star
    markers showing the switch to quiescence mode. The black line is the
    SFH based on abundance matching from
    \citet{Behroozi2013}. The grey band shows the errors on
    individual galaxies.}
  \label{fig:sfh}
\end{figure}


\subsection{Star formation history}
\label{sec:sfh}
The abundance matching technique predicts the expected stellar mass for
a given halo mass at a given time. This
information can be translated into the star formation rate of a galaxy
throughout it's lifetime, it's star formation history (SFH).  The SFH
is constructed by considering the formation time of all stars and
summing their mass in bins of $1\,$Gyr. Figure \ref{fig:sfh} shows the
SFHs of each galaxy. The SF decreases over time after the onset of the
quiescence mode (the black star in each panel). The initial
($1-3$\,Gyr) and the final ($12-13$\,Gyr) times broadly agree with
the abundance matching results \citep[black solid
  line,][]{Behroozi2013}. It should be noted that the SFHs from
abundance matching shown here are for generic galaxies with $z=0$ halo
masses in the range $10^{13}-10^{13.2}\,\Msun$. Our galaxies's final
halo masses all fall within this range.

Due to the swiftness with which the star formation drops at the onset
of the quiescence mode, there is a discrepancy at intermediate
times ($4-11$\,Gyr). 
{Again, we compare Halo3 (blue line) and Halo3late (light blue
dotted line) where the cooling mode is switched later, at
$z=1.5$. The SF drops
to values below $10$\,\Msun yr$^{-1}$ after the switch and then
progesses similarly to Halo3. From figure \ref{fig:ms-mh}, we know the
final stellar mass lies
just outside the $1-\sigma$ range of the abundance matching relation. This shows that some amount of scatter
in $\Moff$ is necessary (and expected) to reproduce the knee of the
relation. 

The intermediate times are not reproduced by these SFHs and
we acknowledge the limitations of the model in this
  respect here, since we are not trying to produce the diversity of
  galaxies and SFHs seen in observations with this simulation. Ours is a
  heuristic quenching model with which we are merely trying to show
  that it can make large ellipticals.}
In the central panels the black dashed line shows results for the
high resolution version of Halo2 (Halo2h); the fiducial and high resolution simulations
agree well.

\begin{figure*}
  \centering
 \includegraphics[width=0.3\textwidth]{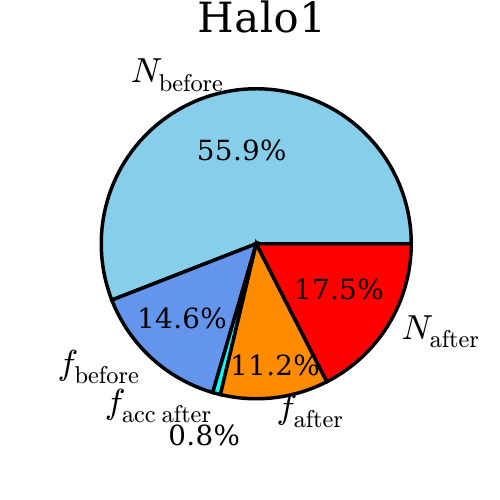}
\includegraphics[width=0.3\textwidth]{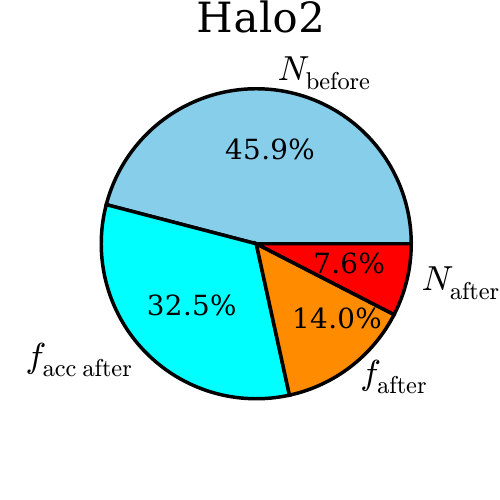}
\includegraphics[width=0.3\textwidth]{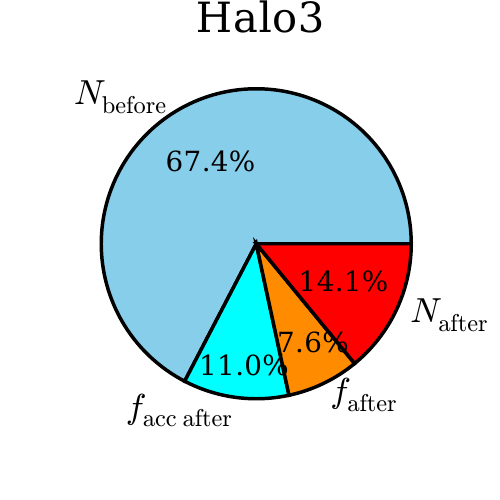}
  \caption{Pie charts for each halo showing the fractional origin of all stars
    present in the halo at $z=0$. $N_{\rm before}$ and $N_{\rm after}$
  are the fraction of stars that formed in the halo before and after the switch in
  cooling mode, respectively. $f_{\rm before}$ is the fraction of
  stars that
formed and were accreted before the switch, while $f_{\rm acc\
  after}$ are the fraction of stars formed before and accreted after
the switch. $f_{\rm after}$ is the fraction formed and accreted
after the switch in cooling modes. Both Halo2 and Halo3 have
negligible fractions of stars that accreted before, $f_{\rm
  before}$, so these were omitted from the plot.}
  \label{fig:pies}
\end{figure*}

Figure \ref{fig:pies} shows the origin of all stars in each halo at
$z=0$. The red and orange wedges are the fraction of stars that formed
(in-situ and ex-situ) after the change in cooling mode. The total
percentages are 28.7\%, 21.6\% and 21.7\% for Halo1, Halo2 and Halo3,
respectively. This means a small, but non-negligible fraction of stars
form at late times after the onset of the quiescence model. The
accreted fraction, or fraction of stars formed ex-situ is 26.6\%,
46.5\% and 18.6\%, respectively, in good agreement with the results
from the Illustris simulation \citep{Rodriguez2016}, in which quenching is achieved via
AGN feedback using both radio and quasar mode.

\begin{figure}
  \centering
 \includegraphics[width=0.47\textwidth]{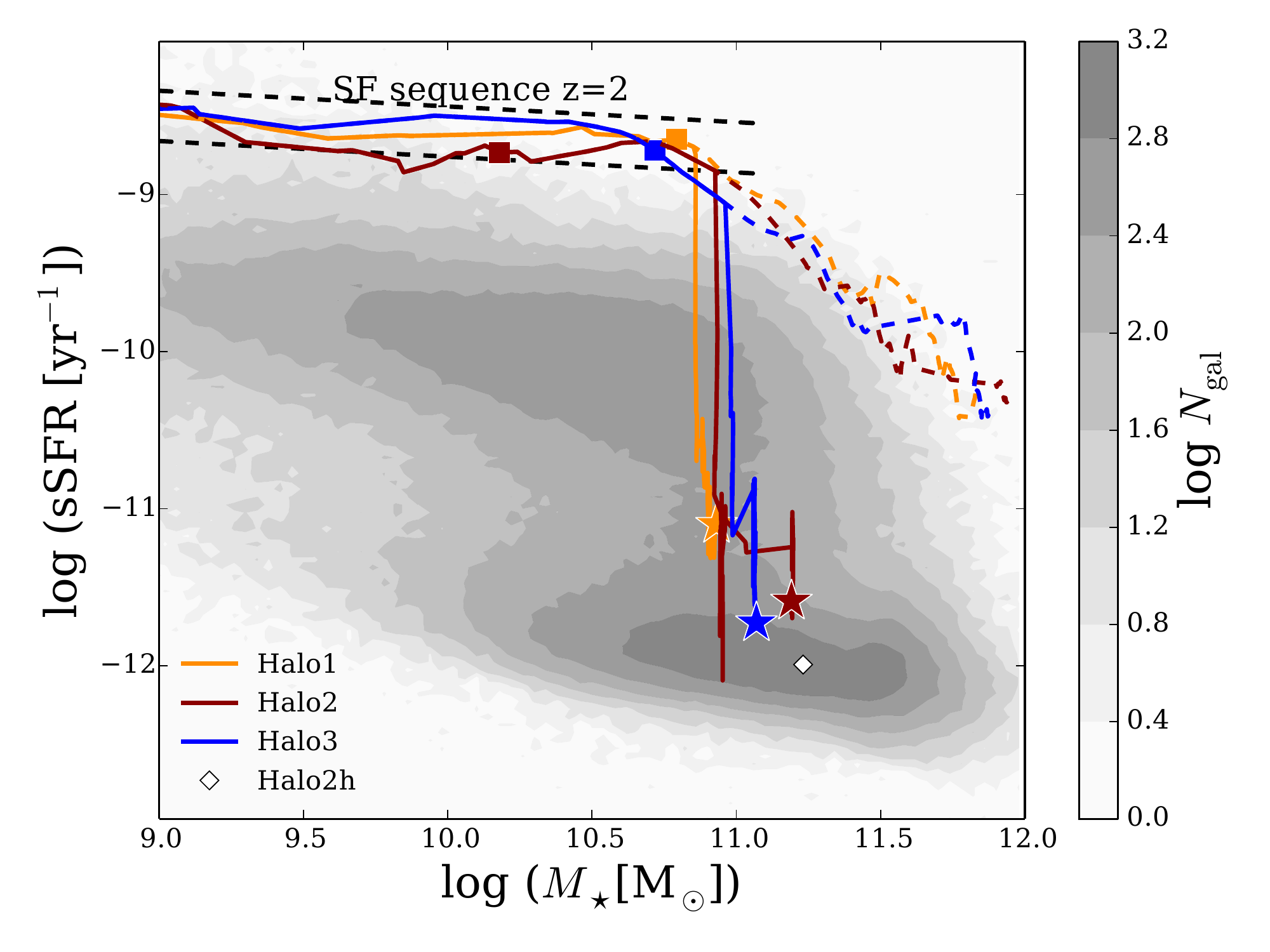}
  \caption{Specific star formation rate (sSFR) versus stellar
    mass. The full lines show the evolution of our galaxies after
    cooling is turned off. Dashed colored lines show the evolution of
    our galaxies if not change is made to the cooling. The evolution
    direction is downwards, where squares show the position at $z=3$
    and stars the position at $z=0$. We plot the $z=0$ position of our
    high resolution version of Halo2 at a white diamond, which can be
    compared to the dark red star. The gray contours are SDSS data
    with $z<0.2$. The stellar masses follow the calculation by
    Kauffmann et al. (2003) and Salim et al.  (2007). The sSFR follow
    Brinchmann et al. (2004). We also plot the extent of the  star
    forming main sequence at $z=2$ from Daddi et al. 2007 as two
    black, dashed lines. }
  \label{fig:ssfr}
\end{figure}

\begin{figure*}
  \centering
 \includegraphics[width=0.9\textwidth]{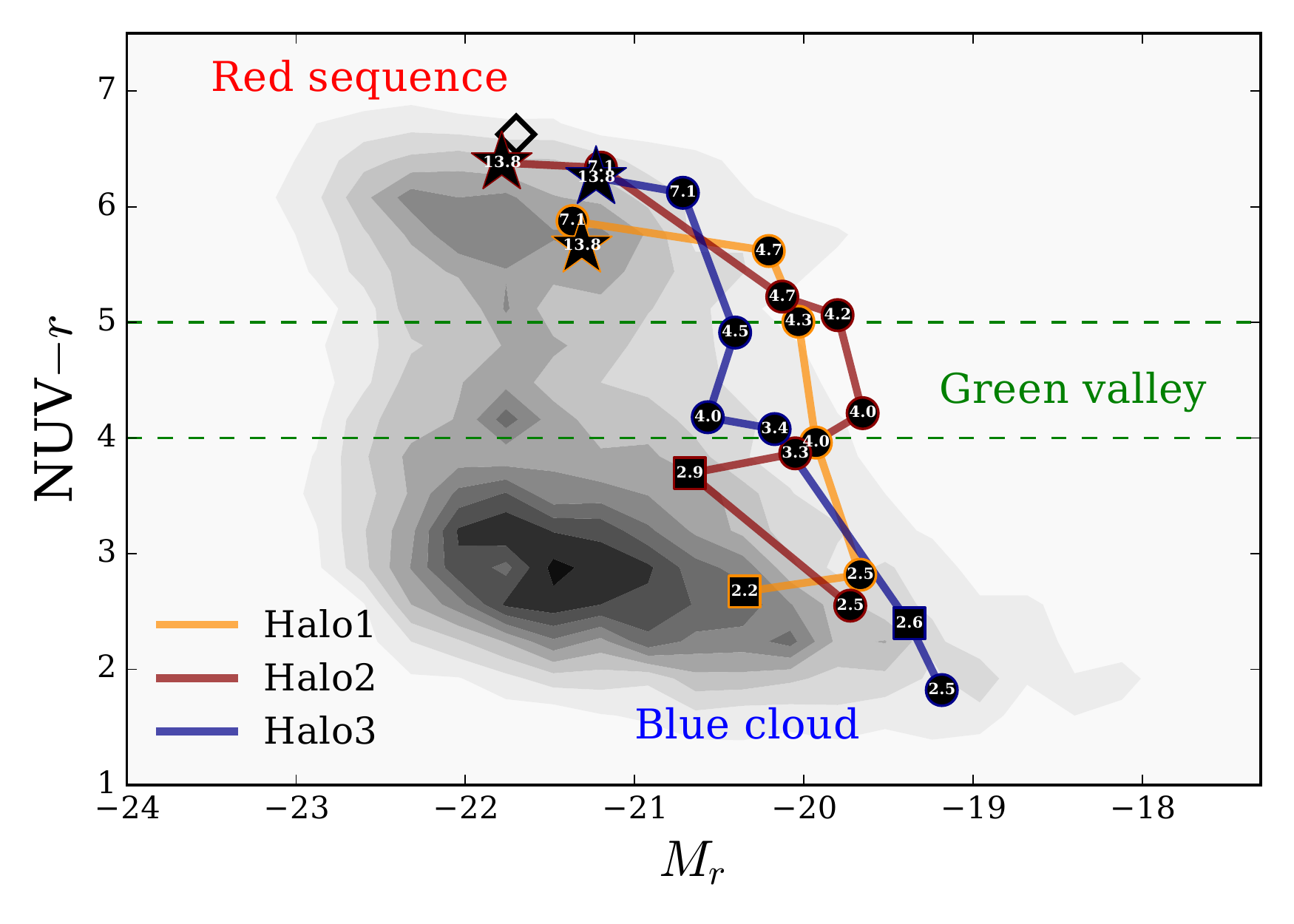}
  \caption{We present the color - magnitude diagram with SDSS and
    GALEX data for $z<0.22$ as in \citet{Salim2007}. We follow our
    galaxy by calculating the magnitudes with the radiative transfer
    code \texttt{SUNRISE}. The black numbers indicate the cosmic time
    in Gigayears at each point. {Square points show the time at which
    the cooling mode is switched.}}
  \label{fig:greenvalley}
\end{figure*}

\subsection{Specific star formation rate}
\label{sec:ssfr}
The specific star formation rate (sSFR) defined as SFR/M$_*$ is a
measure of the efficiency with which a galaxy turns gas into
stars. The bi-modality of ``blue'' and ``red'' galaxies is apparent in
their sSFRs. The distribution of galaxies in the stellar mass - sSFR
plane shows a clear distinction between the ``star forming main
sequence'' of galaxies with sSFR$\sim10^{-10}$yr$^{-1}$ at $z=0$ and a
population of galaxies with higher average stellar masses and
sSFR$\sim10^{-12}$yr$^{-1}$. Figure \ref{fig:ssfr} shows the $z<0.2$
SDSS density of galaxies on this plane as a grey
contours. The stellar masses are taken from \citet{Kauffmann2003} and
\citet{Salim2007} while the sSFRs are from \citet{Brinchmann2004}.

Our galaxies's evolutionary tracks are the coloured lines as detailed
in the legend. Their $z=3$ and $z=0$ positions are marked by squares
and stars, respectively.  The $z=2$ star forming main sequence from
\citet{Daddi2007} is shown as the area between the two black dashed
lines. The tracks evolve on this sequence until the cooling mode is
changed. Then they lose the majority of their active star formation
and arrive at sSFRs of $10^{-11}-10^{-12}$yr$^{-1}$. These values are
in agreement with massive galaxies in the SDSS observations.  The
dashed lines in the respective colours show the evolution of the
galaxies without a switch into quiescence mode. In this case, the
sSFRs decrease more slowly and don't arrive at the low values
necessary to match the SDSS sample.

{We note that models such as \citet{Dutton2015}, although similar
  in other respects, will not be able to reproduce the sSFR as our
  model does. Their model has zero SF after $z=2$ and, thus, predicts no
  absolute value for the sSFR at late times. As shown in figure
  \ref{fig:pies}, of the order of $25\%$ of our final stellar mass is
  formed after the cooling mode switch. This amount of SF is in
  agreement with the SDSS measurements of the present-day sSFR of
  massive ellipticals.}

\subsection{Color magnitude diagram - The green valley}
\label{sec:cmd}

The relatively sparse number of galaxies populating the color
magnitude diagram between the blue cloud and the red sequence,
a.k.a. the green valley, presents a difficulty in understanding the
transition from blue, star forming galaxies to red-and-dead
galaxies. {In this
  work, we will use the term green valley to describe the region of
  the UV-optical diagram between $4<{\rm NUV}-r<5$, following
  \citet{Salim2014}. Other authors have defined it in different ways,
  which we note might lead to variations to the results we present here.} This difficulty lies in the expected time it takes for a
chromatic and morphological transition. 

SDSS/GALEX data ($z<0.22$, see
  \citealt{Salim2007}) obtain an estimate of approximately 1\,Gyr for
the transition. To compare our galaxies's evolutionary paths across
the green valley, we run the radiative transfer code \texttt{SUNRISE}
to take the dust extinction and induced reddening into account. We obtain mock fluxes
of the Sloan $r$-band and the GALEX near ultra-violet (NUV) filter at a
few redshifts. {The derived magnitudes are rest-frame
  magnitudes inside our \texttt{SUNRISE} box of $50\times50$\,kpc.}

The resulting tracks across the CMD are shown in figure
\ref{fig:greenvalley}. The grey contours are the SDSS/GALEX density
map. We follow \citet{Salim2014} and define the green valley between
NUV-$r=4-5$. The lines constitute the tracks of our three galaxies,
while the black dots show the points at which a radiative transfer
calculation was made. The number in the dots denotes the time in
Gigayears at each point. Halo2 (dark red line) transits the green
valley between $t=3.3$\,Gyr and $t=4.2$\,Gyr. Consequently, it
transits the green valley in approximately $0.9$\,Gyr. Halo3 (blue
line) crosses in $\sim1.1$\,Gyr, while Halo1 (dark yellow line) has
an even shorter transit time of $\sim0.3$\,Gyr.  Finally, since we have a finite
number of simulations outputs, we approximate the error on the crossing times with the average time
between two consecutive snapshots, which is around $0.2$\,Gyr. 

{The \texttt{SUNRISE} dust model affects the resulting magnitudes
  most strongly during ongoing SF, because new stars are assumed to
  stay enshrouded in their birth cloud for up to $10$\,Myr. Without SF,
the dust mass scales with the gas metallicity. This effect is
important after the cooling switch, since much gas is initially still
in the central galaxy. With time, the effect decreases since gas, and
hence dust, is expelled due to feedback.}
It is worth noting that our galaxies do not cross the green valley
along the area of highest density in the observations, which would be
around $M_r\sim-22$. Instead, they cross around $M_r\sim-20$ and become
brighter after transiting the green valley. Since there is little SF
at late times, this brightening is caused by the loss of dust in the
galaxy. Decreasing cold gas fractions in the center cause the estimated
dust extinction to decrease when more gas is heated and held in the
hot halo outside of the galaxy.

{We acknowledge that there are many parameters in the
  \texttt{SUNRISE} model that could change the resulting
  magnitudes, i.e. the grid cell size and the dust grain size among others. Due to the simplicity of our starvation model, we did
  not see the need to fine tune these. The important result of this
  exercise is that the relative change in color (here the
  NUV-$r$) happens in less than 1 Gyr. We expect this relative result
  to be less affected by the specific model parameters. 

Moreover, our model is just one quenching scenario to rapidly cross
the green valley. Authors such as \citet{Smethurst2015} and \citet{Trayford2016}
show that different galaxy populations might well quench through a
variety of mechanisms that happen with differing timescales. Some
proposed causes of
reddening and morphological change are secular evolution, major
mergers and AGN.}

\subsection{Sizes and profiles}
\label{sec:rm}

\begin{figure}
  \centering
 \includegraphics[width=0.5\textwidth]{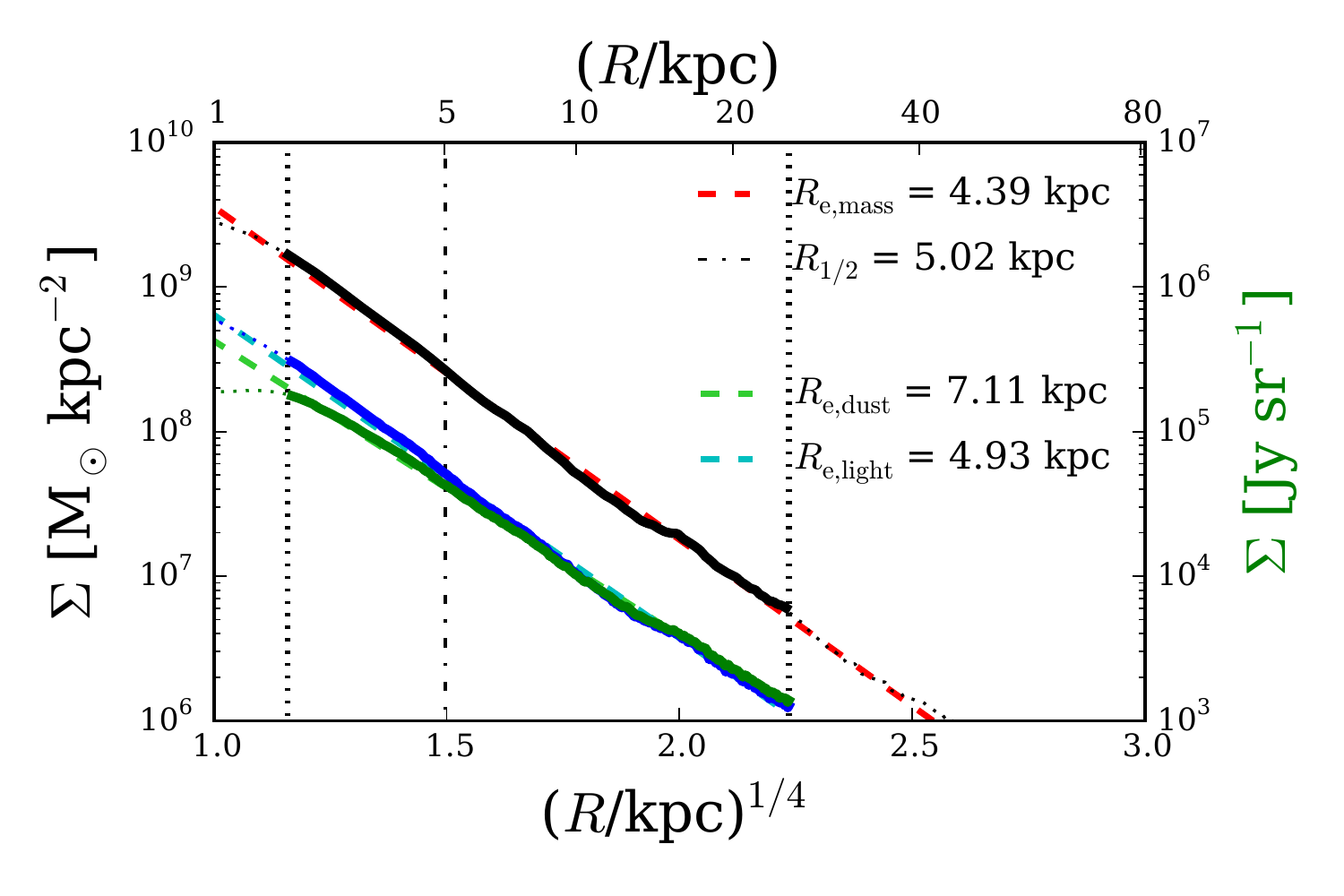}
  \caption{{Surface density (black solid and black dotted line) and
    surface brightness profiles for Halo2h. The surface brightness profiles are derived from
    \texttt{SUNRISE} images with dust (green solid and dotted line)
    and without dust (blue solid and dotted line) and
    scale with the y-axis on the right.  Each profile is fit with a de
    Vaucouleur profile, which are shown a thick dashed lines in red (density),
    cyan (brightness without dust) and light green (brightness with dust). The vertical black dotted
    lines show the range of the fit, which is between three smoothing
    lengths (1.8\,kpc at this resolution) and 25\,kpc, which is the extent of the \texttt{SUNRISE}
    image. The effective radii of the fits are shown in
    the legend. The vertical black dot-dashed line shows the position
    of the half-mass radius derived through integration and its value
    is also shown in the legend.}}
  \label{fig:profile}
\end{figure}

Matching the sizes of elliptical galaxies in simulations has been
notoriously difficult, generally producing too compact centers. This
difficulty might in part be related to the task of measuring the size
in a comparable manner.  {Figure \ref{fig:profile} shows the
\texttt{SUNRISE} surface brightness profile in radial annuli (green
with dust
and blue without). The cyan and light green dashed
lines shows de Vaucouleur (Sersic profile with $n=4$) fit to the
surface brightness profile and the resulting effective radius is shown
in the legend. The fit is performed in the radial range from 3
smoothing lengths ($\epsilon_{\rm gas}\sim 3.3\,$kpc for the fiducial
resolution) out to 25\,kpc. The fitting range is shown between the
vertical dotted lines. We compare the light profile with the
mass profile by projecting the galaxy onto the same axis as the
\texttt{SUNRISE} image, resulting in a surface mass density
profile. The same fit is performed and shown as a red dashed line.
The dot-dashed line shows half-mass radius, $R_{1/2}$, obtained from the 
integrated stellar mass profile as indicated in the legend.}

%

\begin{figure}
  \centering
 \includegraphics[width=0.47\textwidth]{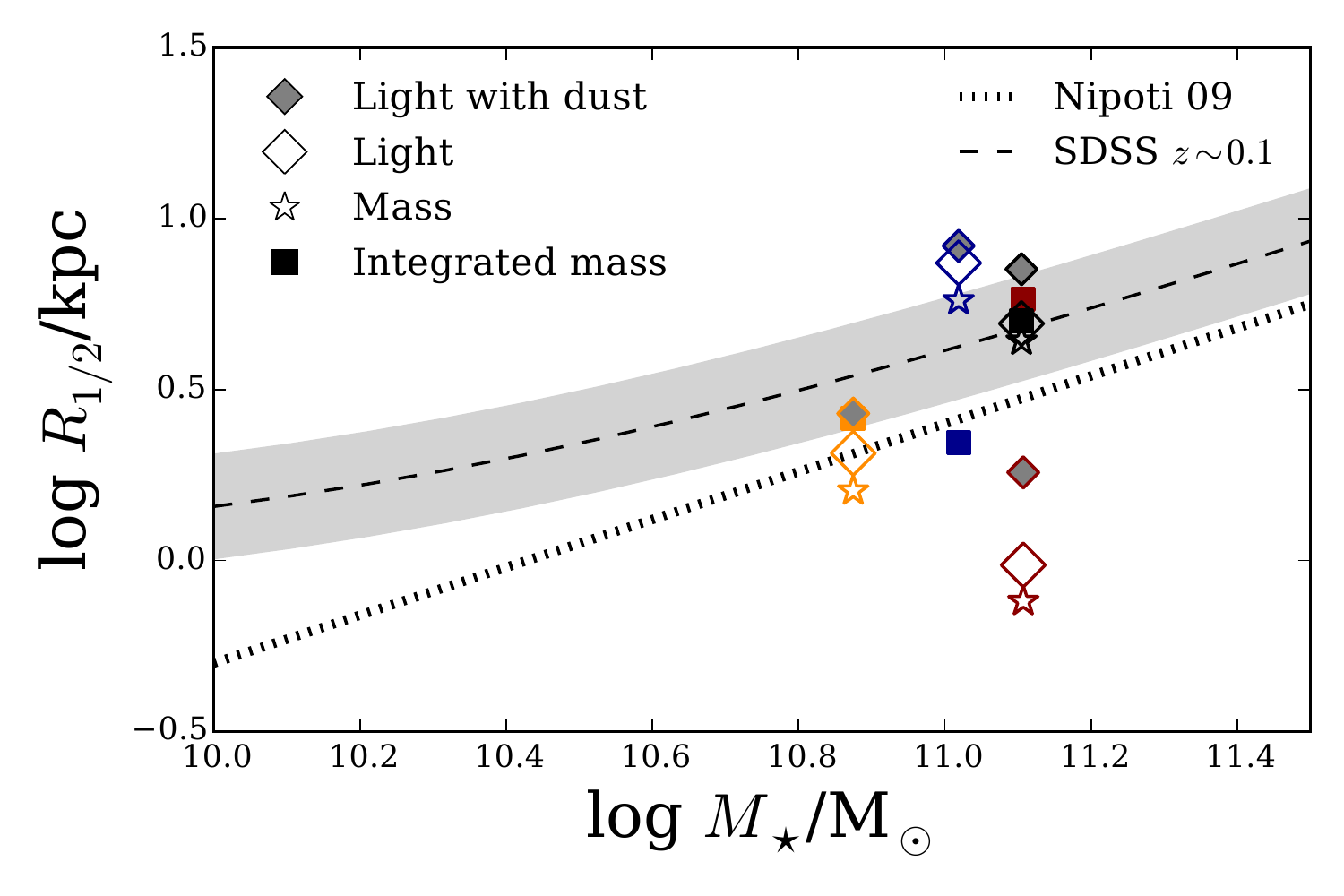}
 \caption{Size-mass relation for our three fiducial galaxies. The
   filled squares show the half-mass radii calculated from the
   integrated mass profile. The diamonds
 show the effective radii derived from the deVaucouleur fits on the
 \texttt{SUNRISE} profiles with and without dust in grey and unfilled,
 respectively. The open stars are the effective radius from the
 deVaucouleur fits to the surface density profiles. Each galaxy is shown it its respectively color (Halo1 in orange, Halo2 in dark red,
 Halo3 in blue). The black symbols indicate
the high resolution galaxy Halo2h. The black dashed line
and grey band are the analytic fit of the $z\sim0.1$ SDSS data by
\citet{Dutton2013}. The black dotted line shows the fit by \citet{Nipoti2009}.}
  \label{fig:size}
\end{figure}

{Figure \ref{fig:size} shows the  $z=0$ sizes of all three galaxies
using all four of the above measures.  The filled squares are the half-mass
sizes when considering the integrated mass profile, $R_{1/2}$.  The gray and open diamonds
are the effective radius from the fit to the surface brightness
profile with and without dust, $R_{\rm e, light}$ and $R_{\rm e, dust}$ respectively. The open star symbols are
the sizes derived from the surface density profiles, $R_{\rm e, mass}$.} We compare
these with the fitting function (black dashed line) from
\citet{Dutton2013} who used SDSS early-type galaxies within the
redshift range $0.005 < z < 0.3$ and a median redshift of $\sim 0.1$
to produce the average as well as the $16^{th}$ and $84^{th}$
percentiles (grey area) of the observational sample. {The observed
relation from \citet{Nipoti2009} who used the SLACS sample of local
early-type galaxies is shown as a black dotted line.}

{The extent of Halo1 and Halo2h are in good agreement with
the observational data. 
Halo3 has a dense stellar core which decreases the integrated mass
profile and is too compact to be consistent with a deVaucouleur in the center.
The profile of Halo2 is different from its high resolution counterpart
due to an incoming merger. We are only able to fit the central part of
the profile ($<10\,$kpc) and, thus, get a steeper slope and smaller
size. The integrated mass profile for both Halo2 (dark red) and Halo2h
(black) are
well within the observational constraints. The similarity of the high
resolution Halo2h with the lower resolution Halo2
sizes gives confidence to the results of our fiducial resolution
simulations. Given
this simple quiescence model, the agreement between the simulation and
observations is encouraging.}

\subsection{Faber-Jackson relation}
\label{sec:vm}

\begin{figure}
  \centering
 \includegraphics[width=0.47\textwidth]{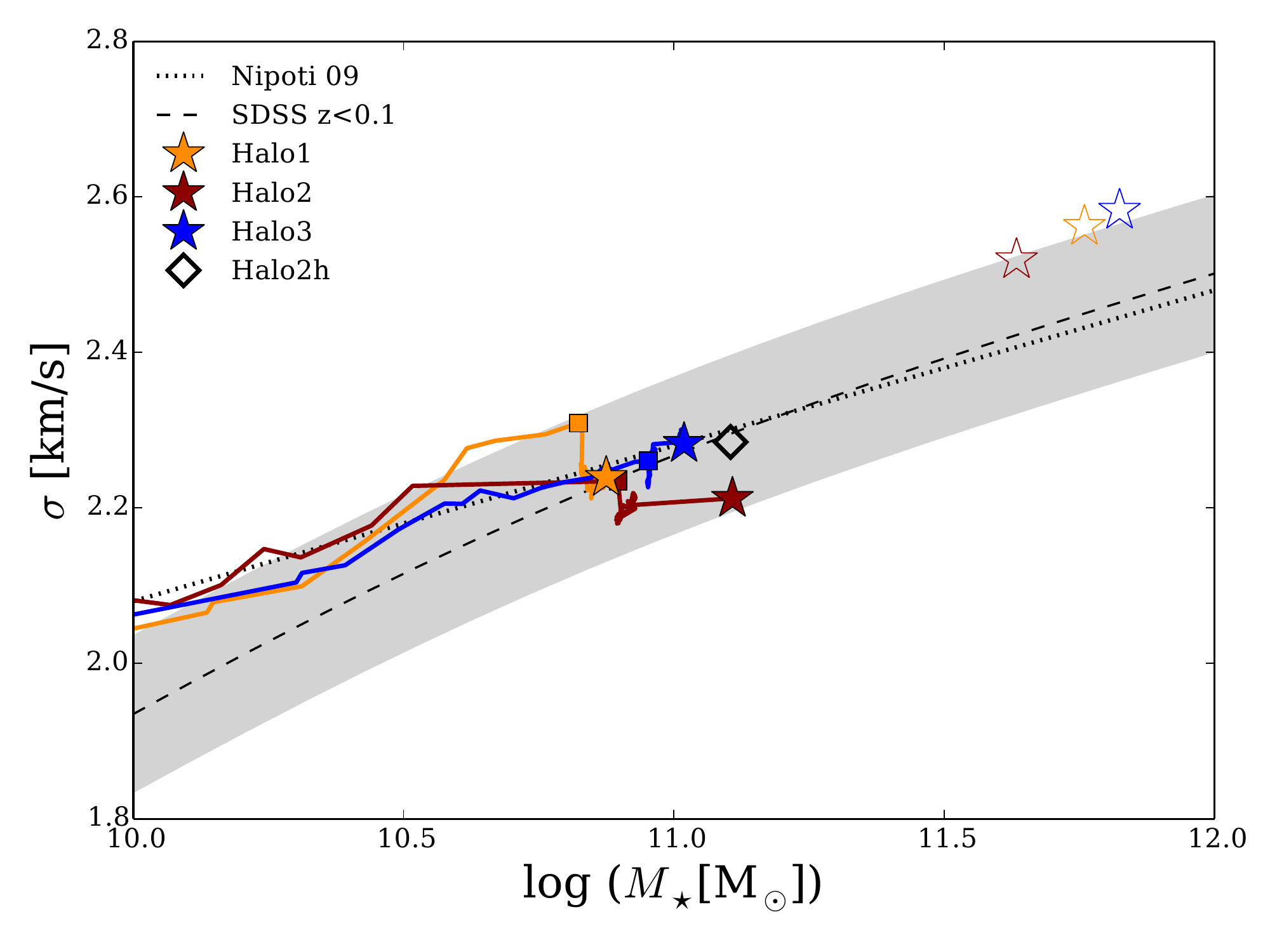}
  \caption{Stellar mass - stellar velocity dispersion with the
    observational data (SDSS z=0.1) fit by Dutton et al. 2013 (black
    dashed line). The 16th and 84th percentiles are shown as a grey
    shaded area. The star markers are our galaxy at z=0. The black
    diamond shows the $z=0$ position of Halo2 in high resolution. Open
  stars show the $z=0$ value for the simulations (in the corresponding color) without changing the
  cooling mode (overcooled). Filled squares indicate the stellar mass
  at the point at which cooling is switched off. The marginal
  evolution after this point suggests that the galaxy was already
  bulge-dominated at the time of \Moff.}
  \label{fig:m-sigma}
\end{figure}

The stellar mass - stellar velocity dispersion relation is one
projection of the fundamental plane of elliptical galaxies. Elliptical
galaxies have on average higher velocity dispersions than star-forming
galaxies. This kinematical change is likely linked to the quenching
mechanism and therefore provides a critical constraint for quenching
models.

Figure \ref{fig:m-sigma} shows where on the relation our three
galaxies fall at $z=0$. Lines in the appropriate color show the tracks
of each galaxy up to $z=0$. To portray the observational relation, we
show the analytical fit to SDSS early-type galaxies ($z<0.1$) from
\citet{Dutton2013} as a black dashed line {and the scaling
  relation derived by \citet{Nipoti2009} (using SLACS early-type
  galaxies at $z=0$) as a black dotted line.}

Both Halo1 and Halo3 fall directly on the mean today, while Halo2
lies slight below but still inside the scatter. The high resolution
version of Halo2 sits on the mean. The simulations without a change
in the cooling mode produce $z=0$ velocity dispersion values too high
for their given stellar mass, as shown with the open star symbols. This is corrected with our simple
quiescence model.

\section{SUMMARY}
\label{sec:summary}

In this paper we have critically examined the properties exhibited by
massive galaxies whose star formation is quenched entirely by
starvation. The starvation is initiated by means of a simple
quiescence model in which all halo gas above a threshold temperature
is prevented from cooling. This is a simple way to simulate the
effects of star formation fading away. Existing stars drive winds and
end as supernovae, removing cold gas from the galaxy. The cold gas
reservoir is depleted over time because it is: 1) turned into stars or
2) heated. Each cycle of star
formation heats a varying fraction of the star forming gas to above
the temperature threshold. The threshold temperature, \Toff, and halo mass at
which the switch is initiated, \Moff, are chosen to match the halo mass -
stellar mass relation derived by abundance matching.

Over time, this causes less and less gas to
be available for future cycles of star formation. Since we switch to
this cooling mode at the height of SF (SFR$\sim10^2\,\Msun$yr$^{-1}$),
the initial effect is large and SF drops quickly
(to SFR$\sim10\,\Msun$yr$^{-1}$ in $\sim1$\,Gyr). Later, the SFR
slowly settles down to a value of $\sim1\,\Msun$yr$^{-1}$.

The resulting galaxy properties are compared to an array of
observations. The SFRs follow the downward trend after the height of
SF at around $z=2.5$. But they are about 1\,dex too low at intermediate
times ($z=2-0.5$). We vary the parameter $M_{\rm off}$ to show that a
galaxy-to-galaxy scatter in the quenching time may smooth the knee of
the SFH derived from abundance matching. Since we do not include other galaxy
types, we don't expect it to be able to represent the overall envelope.
The sSFR evolution with increasing stellar mass initially follows the
$z=2$ SF main sequence and then drops to low values
(sSFR$\sim10^{-12}\,{\rm yr}^{-1}$) in accordance with the SDSS
values for the quenched population at $z=0$.

A key insight of this paper is that allowing star formation to fade by
starvation can produce green valley crossing times of well below
1\,Gyr, contrary to common assumptions of the opposite. We were able
to show this by performing radiative transfer post-processing on the
evolving galaxies and estimating their mock $r$-band and $NUV$
fluxes. These were used to construct evolutionary tracks across the
CMD and estimating crossing times. The three galaxies presented
in this work deliver crossing times in the range $0.3-1.1\,$Gyr. The
swiftness of the chromatic transformation is surprising and we
conclude that the quenching process does not require a single event
that instantaneously rids the galaxy of its gas reservoir. Rather a
gradual process in which the gas is heated and made unavailable for SF
is sufficient to produce the required transformation. Although we
  note that this result is dependent on our quenching model, and also
  on the dust modelling in \texttt{SUNRISE}.

Additionally, we compare the mass profiles and resulting sizes of our
galaxies with the observed elliptical population at $z=0$. These match
surprisingly well. The two galaxies with smaller stellar masses might be
slightly too compact, depending on the method of calculating the
sizes. {Dust affects the sizes to some degree, making them
  slightly larger due to the decreased brightness in the central
  regions of the galaxies.} Also the velocity dispersion at $z=0$ matches observations,
signifying that the stellar kinematics correctly change with the
morphological transformation. 

We recognise that this simple model is too simple to delve further
into the properties of these elliptical galaxies. It is
necessary to move towards more physically motivated models of
quenching in future work to be able to further understand the
quenching process. 


\section*{Acknowledgments} 
We thank the anonymous referee for very helpful and constructive comments.
We thank Aura Obreja and Glenn van de Ven
for helpful suggestions and useful conversations.
The simulations were run using the galaxy formation code
\texttt{GASOLINE2}, developed and written by Tom Quinn and James
Wadsley.  Without their contributions, this paper would have been
impossible.
TAG and AVM acknowledge funding by Sonderforschungsbereich SFB
881 ``The Milky Way System'' (subproject A1) of the German Research
Foundation (DFG).
The analysis was performed using the pynbody package
\citep[http://pynbody.github.io/,][]{Pontzen2013}, written by Andrew
Pontzen and Rok Ro\v{s}kar. 
This research was carried out on the High Performance Computing resources at New York University Abu Dhabi;
on  the {\sc theo} cluster of the
Max-Planck-Institut f\"{u}r Astronomie and the {\sc hydra} cluster at
the Rechenzentrum in Garching through the Collaborative Research
Center (SFB 881) ``The Milky Way System'' (subproject Z2),  hosted and
co-funded by the J\"{u}lich Supercomputing Center (JSC).
We are happy to use pie charts in a scientific publication that show
something other than the cosmological distribution of matter and
energy in the Universe.

\balance
\bibliographystyle{mnras}
\bibliography{bib}

\appendix

\bsp

\label{lastpage}

\end{document}